\documentclass[twocolumn,preprintnumbers,amsmath,amssymb,prl]{revtex4}

\usepackage{graphicx}
\usepackage{dcolumn}
\usepackage{bm}
\usepackage{amssymb}
\usepackage{epstopdf}
\usepackage{color}

\begin{document}

\title{Optically induced hybrid Bose-Fermi system in quantum wells with different charge carriers}

\author{O. V. Kibis}
\author{M. V. Boev}\email{Oleg.Kibis(c)nstu.ru}
\author{V. M. Kovalev}
\affiliation {Department of Applied and Theoretical Physics, Novosibirsk~State~Technical~University,
Karl~Marx~Avenue~20,~Novosibirsk~630073,~Russia}

\begin{abstract}
It is demonstrated theoretically that the circularly polarized irradiation of two-dimensional conducting systems can produce the composite bosons consisting of two electrons with different effective masses (different charge carriers), which are stable due to the Fermi sea of conduction electrons. As a result, the optically induced mixture of paired electrons and normal conduction electrons (the hybrid Bose-Fermi system) appears. Elementary excitations in such a hybrid system are analyzed and possible manifestations of the light-induced electron pairing are discussed for semiconductor quantum wells.
\end{abstract}

\maketitle

{\it Introduction.} The modification of electronic properties of various quantum systems by an off-resonant high-frequency electromagnetic field (the Floquet engineering) became the
established research area of modern physics, which resulted in many fundamental effects studied both theoretically and experimentally ~\cite{Oka_2019,Bukov_2015,Lindner_2011,Savenko_2012,Kozin_2018,Rechtsman_2013}.
Recently, we demonstrated that a high-frequency circularly polarized electromagnetic field crucially modifies the Coulomb interaction in two-dimensional (2D) systems, inducing the attractive area in the core of the repulsive Coulomb potential~\cite{Kibis_2019}. As a consequence, the quasi-stationary electron states bound at repulsive scatterers~\cite{Kibis_2020,Iorsh_2020} and, particularly, the pairing of electrons with different effective masses~\cite{Kibis_2019} can appear. However, the single electron pair proved to be instable with very short lifetime that prevents to observe experimentally effects arisen from the pairing. In the present Letter, we developed the many-body theory of the light-induced electron pairing in semiconductor quantum wells and found that the Fermi sea of conduction electrons stabilize the pairs. As a consequence, the hybrid Bose-Fermi system consisting of the paired electrons and normal conduction electrons appears. In what follows, the physical properties of such a system are analyzed and the optically induced phenomena originated from the Bose nature of paired electrons are discussed.

{\it Model.} Let us consider a two-dimensional (2D) nanostructure containing charge carriers with different effective masses, where the energy spectrum of the two ground subbands is $\varepsilon_{1}(\mathbf{k})=-\Delta_0/2+\hbar^2k^2/2m_{1}$ and $\varepsilon_{2}(\mathbf{k})=\Delta_0/2+\hbar^2k^2/2m_{2}$, $\Delta_0$ is the energy distance between the subbands, $\mathbf{k}=(k_x,k_y)$ is the momentum of charge carrier in the 2D plane, and $m_{1,2}$ are the effective masses in the subbands. For definiteness, we will develop the theory for conduction electrons, which can be easily generalized for 2D hole nanostructures. In the presence of a circularly polarized electromagnetic wave incident normally to the 2D structure, the Coulomb interaction of two electrons from the subbands $\varepsilon_{1}(\mathbf{k})$ and $\varepsilon_{2}(\mathbf{k})$ is described by the Hamiltonian $\hat{\cal H}_{12}(t)=\hat{\cal H}_{1}+\hat{\cal H}_{2}+U(\mathbf{r}_1-\mathbf{r}_2)$, where $\hat{\cal H}_{1,2}=(\hat{\mathbf{p}}_{1,2}-e\mathbf{A}(t)/c)^2/2m_{1,2}$ are the Hamiltonians of free conduction electrons irradiated by the wave, $\hat{\mathbf{p}}_{1,2}=-i\hbar\partial/\partial\mathbf{r}_{1,2}$ are the plane momentum operators of the electrons, $\mathbf{r}_{1,2}$ are the plane radius vectors of the electrons,
\begin{equation}\label{A}
\mathbf{A}(t)=(A_x,A_y)=[cE_0/\omega_0](\sin\omega_0 t,\,\cos\omega_0
t)
\end{equation}
is the vector potential of the wave, $E_{0}$ is the electric field amplitude of the
wave, $\omega_0$ is the wave frequency, $U(\mathbf{r}_{12})=e^2/{r}_{12}$ is the potential energy of the repulsive Coulomb interaction between the electrons, and $\mathbf{r}_{12}=\mathbf{r}_{1}-\mathbf{r}_{2}$ is the radius vector of relative position of the electrons. If the field frequency $\omega_0$ is high enough and lies far from
characteristic resonant electron frequencies, the interaction between the field (\ref{A}) and electrons can be considered within the conventional Floquet theory. As a result, the time-dependent Hamiltonian $\hat{\cal H}_{12}(t)$ can be reduced to the effective stationary Hamiltonian~\cite{Kibis_2019,Kibis_2020},
\begin{equation}\label{H01}
\hat{\cal H}_0=\frac{\hat{\mathbf p}_1^2}{2m_1}+\frac{\hat{\mathbf p}_2^2}{2m_2}+U_0(r_{12}),
\end{equation}
where
\begin{eqnarray}\label{U0}
U_0({r}_{12})&=&\frac{1}{2\pi}\int_{-\pi}^{\pi}U_{12}\big(\mathbf{r}_{12}-\mathbf{r}_0(t)\big)\,d(\omega_0
t)\nonumber\\
&=&\left\{\begin{array}{rl}
({2e^2}/{\pi r_0})K\left({r_{12}}/{r_0}\right),
&{r_{12}}/{r_0}\leq1\\\\
({2e^2}/{\pi r_{12}})K\left({r_0}/{r_{12}}\right),
&{r_{12}}/{r_0}\geq1
\end{array}\right.
\end{eqnarray}
is the repulsive Coulomb potential dressed by the circularly polarized field (\ref{A}) (dressed potential), $K(\xi)$ is the
complete elliptical integral of the first kind, $\mathbf{r}_0(t)=(-r_0\cos\omega_0 t,\,r_0\sin\omega_0 t)$ is the radius vector describing the classical circular trajectory of a free particle with the charge $e$ and the mass $\overline{m}=m_1m_2/(m_1-m_2)$ in the circularly polarized field (\ref{A}), and $r_0={|e\overline{m}|E_0}/{\omega^2_0}$ is the radius of the trajectory.
\begin{figure}[!ht]
\includegraphics[width=1.\columnwidth]{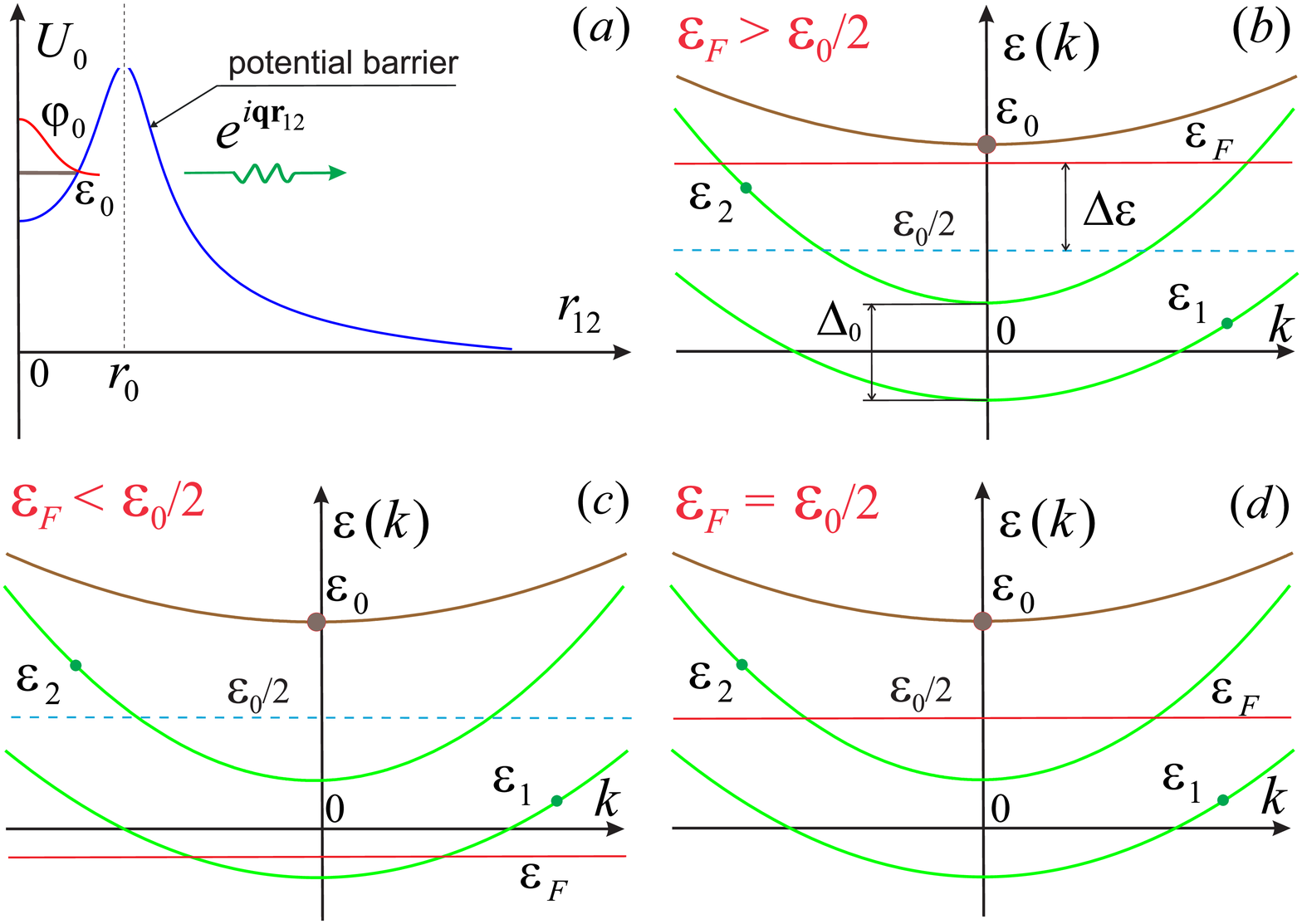}
\caption{Electronic energy structure of the system under consideration: (a)  The light-induced effective potential of electron-electron interaction, $U_0$, with the local minimum which confines the electron pair with the wave function $\varphi_0$ (the red curve) and the energy $\varepsilon_0$ (the brown horizontal line). Normally, the electron pair is instable because of the tunneling from the local minimum through the potential barrier into the states of decoupled conduction electrons with the plane electron wave $e^{i\mathbf{qr}_{12}}$ (the green wave arrow); (b)--(d) The energy spectrum of the subbands of conduction electrons $\varepsilon_{1,2}(k)$ (the two lowest green curves) and the coupled electrons $\varepsilon_{0}(k)$ (the upper brown curve) for different Fermi energies, $\varepsilon_F$. The large brown circle marks the ground  state of coupled electrons with the energy $\varepsilon_0$, whereas the two small green circles mark the states of decoupled electrons with the energies $\varepsilon_{1,2}$ satifying the condition $\varepsilon_1+\varepsilon_2=\varepsilon_0$.}\label{Fig.1}
\end{figure}

Introducing the radius vector of the center of mass of the electrons, ${\mathbf{R}}_{12}=(m_1\mathbf{r}_1+m_2\mathbf{r}_2)/(m_1+m_2)$, the effective Hamiltonian (\ref{H01}) can be rewritten as
\begin{equation}\label{H0}
\hat{\cal H}_0=-\frac{\hbar^2}{2(m_1+m_2)}\frac{\partial^2}{\partial\mathbf{R}_{12}^2}
-\frac{\hbar^2}{2m}\frac{\partial^2}{\partial\mathbf{r}_{12}^2}+
U_0({r}_{12}),
\end{equation}
where $m=m_1m_2/(m_1+m_2)$ is the reduced effective mass.
Since the dressed potential (\ref{U0}) has the local minimum at $\mathbf{r}_{12}=0$ (see Fig.~1a), the solution of the Schr\"odinger problem with the Hamiltonian (\ref{H0}) results in the wave function of coupled electrons $\varphi_\mathbf{K}=(1/\sqrt{S})e^{i\mathbf{K}\mathbf{R}_{12}}\varphi_0(\mathbf{r}_{12})$ and their energy spectrum $\varepsilon_0(\mathbf{K})=\varepsilon_0+\hbar^2K^2/2(m_1+m_2)$, where $S$ is the area of the 2D structure, $\mathbf{K}$ is the momentum of the center of mass of the electron pair, $\varepsilon_0$ is the ground energy of the pair and $\varphi_0(\mathbf{r}_{12})$ is the wave function describing relative motion of the paired electrons with the characteristic radius of the pair $r_0$. It should be noted that the energy of coupled electrons found from the spinless Hamiltonian (\ref{H0}) corresponds to the degenerate singlet and triplet spin states. Although the spin-spin interaction of coupled electrons lifts this degeneracy, the corresponding spin splitting, $\Delta_s\sim(e\hbar/m_ec)^2/r_0^3$, is relativistically small and can be omitted as a first approximation. Since the light-induced electron pairs have the integer spin, they can be considered as composite bosons. It should be stressed also that the potential~(\ref{U0}) turns into the usual repulsive Coulomb potential $U({r}_{12})=e^2/r_{12}$ for $m_1=m_2$ that leads to disappearance of the pairing for electrons with equal effective masses.

It should be noted that other photon-assisted mechanisms of electron pairing are also known (see, e.g., Refs.~\cite{Kavokin_2005,Kroo_2014} and references therein). However, in addition to photons, they are mediated by various elementary excitations (plasmons, polaritons, phonons), whereas the present mechanism is based only on the interaction between electrons and a high-frequency electromagnetic field. Physically, the discussed field-induced attraction between electrons originates from oscillating movement of the electrons in the field. If effective masses of electrons are different, $m_1\neq m_2$, such a movement changes the inter-electron distance harmonically. Therefore, the repulsive Coulomb interaction of them also oscillates harmonically. Averaging the oscillating repulsive Coulomb potential over the oscillation period, we arrive at the effective stationary potential of electron-electron interaction (\ref{U0}), which has the local minimum. Thus, the domain of attraction appears in the core of the repulsive potential~\cite{Kibis_2019}. Regarding to circular polarization of the field (\ref{A}), it is important for the pairing since a linearly polarized field induces the effective potential with a saddle point which cannot hold two interacting electrons (see Fig. 2 in Ref.~\cite{Kibis_2019} and the discussion therein).

The state of coupled electrons with the energy $\varepsilon_0$ is immersed into the continuum of normal electron states of the subbands $\varepsilon_{1}(\mathbf{k})$ and $\varepsilon_{2}(\mathbf{k})$ and separated from them by the potential barrier pictured in Fig.~1a, which confines the wave function $\varphi_0(\mathbf{r}_{12})$ near the local minimum of the dressed potential $U_0({r}_{12})$. Therefore, the coupled electron pair is quasi-stationary and can decay by tunneling electrons through this barrier into the normal electron states of the subbands. The energy of the two decoupled electrons can be written as ${\varepsilon}(\mathbf{q},\mathbf{K})=\hbar^2k_1^2/2m_1+\hbar^2k_2^2/2m_2=\hbar^2q^2/2m+\hbar^2K^2/2(m_1+m_2)$, where $\mathbf{k}_1=m_1\mathbf{K}/(m_1+m_2)+\mathbf{q}$ and $\mathbf{k}_2=m_2\mathbf{K}/(m_1+m_2)-\mathbf{q}$ are the momenta of the two decoupled electrons in the subbands $\varepsilon_{1}(\mathbf{k})$ and $\varepsilon_{2}(\mathbf{k})$, respectively, $\mathbf{q}$ is the momentum of relative motion of the decoupled electrons, and the corresponding electron wave function of the decoupled electrons reads $\psi_{\mathbf{q},\mathbf{K}}=(1/S)e^{i\mathbf{q}\mathbf{r}_{12}}e^{i\mathbf{K}\mathbf{R}_{12}}$.
Because of the translational invariance of the system, the dressed potential (\ref{U0}) does not depend on the center-of-mass position $\mathbf{R}_{12}$ and, therefore, it does not change the total momentum of the pair $\mathbf{K}$. As a consequence, the tunnel decay of the electron pair corresponds to the transition $\varphi_\mathbf{K}\rightarrow\psi_{\mathbf{q},\mathbf{K}}$, which depends only on the relative momentum $\mathbf{q}$ and is pictured schematically in Fig.~1a. The probability of this transition can be easily found under the condition of weak tunneling, $\Gamma_0\ll\varepsilon_0-U_0(0)$, and reads
\begin{equation}\label{Wks}
W_{\mathbf{q}}=\frac{\hbar^2\Gamma_0}{Sm[(\varepsilon_q-\varepsilon_{0})^2+(\Gamma_0/2)^2]},
\end{equation}
where $\Gamma_0$ is the tunneling-induced broadening of the energy level $\varepsilon_0$ and $\varepsilon_q=\hbar^2q^2/2m$ is the kinetic energy of relative motion of the decoupled electrons (see Supplement 1 for details). The energy $\varepsilon_0$ and the energy broadening $\Gamma_0$, which define the probability (\ref{Wks}), can be found numerically from the Schr\"odinger equation with the Hamiltonian (\ref{H0}) within the conventional approach~\cite{Landau_3} and are plotted in Fig.~2 as functions of the dimensionless reduced effective mass $m/m_e=m_1m_2/m_e(m_1+m_2)$ and the characteristic radius of the electron pair $r_0={|e\overline{m}|E_0}/{\omega^2_0}$, where $m_e$ is the electron mass in vacuum. Since the radius $r_0$ depends on both the field amplitude $E_0$ and the field frequency $\omega_0$, the plots pictured in Fig.~2 describe also the effect of the irradiation on the pairs. It follows from Fig. 2b that the broadening $\Gamma_0$  increases with decreasing field amplitude $E_0$. Since an electron pair can exist if its characteristic binding energy exceeds the energy broadening $\Gamma_0$, the considered effect takes place if the intensity of light is large enough.
\begin{figure}[!ht]
\includegraphics[width=1.\columnwidth]{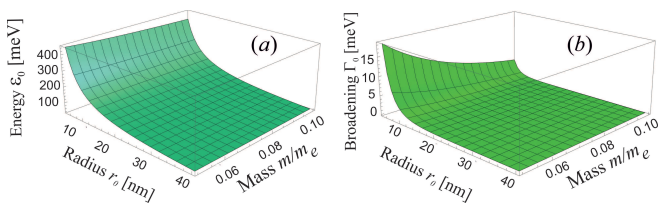}
\caption{Dependence of the electron pair energy $\varepsilon_0$ (a) and the energy broadening $\Gamma_0$ (b) on the dimensionless reduced electron mass $m/m_e=m_1m_2/m_e(m_1+m_2)$ and the characteristic radius of the electron pair $r_0={|e\overline{m}|E_0}/{\omega^2_0}$.}\label{Fig.2}
\end{figure}

Direct summation of the probabilities~(\ref{Wks}) results in the unit total probability of the decay of the coupled electrons, $\sum_\mathbf{q}W_{\mathbf{q}}=1$, i.e. the single electron pair is instable. To consider the effect of degenerate gas of conduction electrons on the pair, it should be stressed first of all that the wave functions of coupled electrons, $\varphi_\mathbf{K}$, and normal conduction electrons, $\psi_{\mathbf{q},\mathbf{K}}$, are separated by the potential barrier pictured in Fig.~1a and, correspondingly, overlap of them is negligible small under the used condition $\Gamma_0\ll\varepsilon_0-U_0(0)$. Therefore, the Pauli principle does not prevent the coexistence of coupled electrons and normal electrons in the same area of space. However, the Pauli principle crucially effects on stability of the pair since it forbids the tunnel decay (decoupling) of the two coupled electrons into the states occupied by normal electrons. As a consequence, the total probability of the decay of the electron pair with the total momentum $\mathbf{K}$ in the presence of normal conduction electrons reads
$W^-_{\mathbf{K}}=\sum_\mathbf{q}W_{\mathbf{q}}(1-f_{\mathbf{k}_1})(1-f_{\mathbf{k}_2})$,
where $f_{\mathbf{k}_{1,2}}=1/[\exp[(\varepsilon_{1,2}(\mathbf{k}_{1,2})-\varepsilon_F)/T+1]$ are the Fermi-Dirac distribution functions for the states of normal electrons with the momenta $\mathbf{k}_1=m_1\mathbf{K}/(m_1+m_2)+\mathbf{q}$ and $\mathbf{k}_2=m_2\mathbf{K}/(m_1+m_2)-\mathbf{q}$ in the subbands $\varepsilon_{1,2}(\mathbf{k})$, $\varepsilon_F$ is the Fermi energy for conduction electrons in these subbands, and $T$ is the temperature. Since the Pauli principle decreases the number of normal electron states which are accessible for the decoupled electrons in the decay process, the decay probability decreases as well, i.e. $W^-_{\mathbf{K}}<1$. Thus, the Fermi gas of normal electrons acts towards to stabilize the coupled electrons. Moreover, due to the Fermi gas, the process of production of coupled electrons from normal electrons (which is inverse relative to the considered decay process) appears. Since the decay and production processes are reversible, their probabilities are described by the same Eq.~(\ref{Wks}). Therefore, the total probability of  production of the coupled electrons from normal electrons is
$W^+_\mathbf{K}=\sum_\mathbf{q}W_{\mathbf{q}}f_{\mathbf{k}_1}f_{\mathbf{k}_2}$.
As a result, the production of electron pairs from the Fermi gas of normal electrons takes place under the condition $W^+_0>W^-_0$, where
\begin{eqnarray}\label{Wp}
W^{\pm}_0&=&\sum_\mathbf{q}
{W_\mathbf{q}}\{\exp[\pm\varepsilon_{1}(\mathbf{q})\mp\varepsilon_F]/T+1\}^{-1}\nonumber\\
&\times&\{\exp[\pm\varepsilon_{2}(\mathbf{q})\mp\varepsilon_F)]/T+1\}^{-1}
\end{eqnarray}
are the probabilities of production ($W^+_0$) and decay ($W^-_0$) of the electron pair in the ground state.

{\it Results and discussion.} For definiteness, we will restrict the following analysis by the case of infinitesimal broadening $\Gamma_0$ and temperature $T$, when the probability (\ref{Wks}) has the delta-function singularity, $W_\mathbf{q}=(2\pi\hbar^2/Sm)\delta(\varepsilon_q-\varepsilon_0)$, and the thermal fluctuations can be neglected. Then the ground coupled state with the energy $\varepsilon_0$ dissociates (arises) to (from) the two states of normal electrons in the subbands $\varepsilon_{1,2}(\mathbf{k})$ with the momenta $\mathbf{k}_1=\mathbf{q}$ and $\mathbf{k}_2=-\mathbf{q}$ corresponding to the electron energies $\varepsilon_1=\varepsilon_1(\mathbf{q})=-\Delta_0/2+\varepsilon_0m_2/(m_1+m_2)$ and $\varepsilon_2=\varepsilon_2(-\mathbf{q})=\Delta_0/2+\varepsilon_0m_1/(m_1+m_2)$, respectively, which are marked by the green circles in Fig.~1(b)--(d) and satisfy the condition $\varepsilon_1+\varepsilon_2=\varepsilon_0$. The direct calculation with using Eqs.~(\ref{Wks})--(\ref{Wp}) results in the inequalities $W^+_0>W^-_0$ for $\varepsilon_F>\varepsilon_0/2$ and $W^+_0<W^-_0$ for $\varepsilon_F<\varepsilon_0/2$. This means that the normal conduction electrons are instable with respect to their coupling for $\varepsilon_F>\varepsilon_0/2$ (see Fig.~1b) and the coupled electrons are instable with respect to their decoupling for $\varepsilon_F<\varepsilon_0/2$ (see Fig.~1c). Correspondingly, the subsystems of normal electrons and coupled electrons are in equilibrium ($W^+_0=W^-_0$) for $\varepsilon_F=\varepsilon_0/2$ (see Fig.~1d), that is the stability condition for the system as a whole. As a consequence, the lifetime of electron pairs is infinite under this condition. It should be noted also that the pair dissociation due to the scattering of them is inelastic process which needs energy to transfer the paired electrons from the bound state with the energy $\varepsilon_0$  to the overlying states of decoupled electrons. Therefore, it can be neglected for the considered limiting case of zero temperature.

To describe the production of coupled electrons under the condition $\varepsilon_F>\varepsilon_0/2$, let us neglect the interaction between the electron pairs, considering them as an ideal Bose gas with the Bose-Einstein distribution function, ${p}_\mathbf{K}=1/[\exp(\varepsilon_{0}(\mathbf{K})-\mu)/T-1]$, where ${p}_\mathbf{K}$ is the averaged density of the pairs, and $\mu$ is their chemical potential. Then the processes of decay and production of electron pairs can be considered formally as a chemical-like reaction in the mixture of the Fermi gas of normal electrons and the Bose gas of paired electrons. Applying the known condition of chemical equilibrium~\cite{Landau_5} to the considered system, we arrive at the equilibrium condition of the mix, $\varepsilon_{F0}=\mu_0/2$, where $\varepsilon_{F0}$ is the equilibrium Fermi energy of the subsystem of normal electrons, $\mu_0$ is the equilibrium chemical potential of the subsystem of coupled electrons, and the factor $1/2$ arises from the two-electronic structure of the Bose molecules. Since the ideal Bose gas is in the Bose-Einstein condensate for $T=0$, the equality $\mu_0=\varepsilon_0$ takes place. Therefore, the condition of thermodynamical equilibrium of the mixture is $\varepsilon_{F0}=\varepsilon_0/2$ and coincides with the condition of equilibrium between the processes of decay and production of the coupled electrons, $W^+_0=W^-_0$, considered above. Since the production of coupled electrons is accompanied by decreasing the Fermi energy of normal electrons, the nonequilibrium state of the mixture with the Fermi energy of normal electrons $\varepsilon_F>\varepsilon_0/2$ (see Fig.~1b) changes towards the equilibrium state with the Fermi energy $\varepsilon_{F0}=\varepsilon_0/2$ (see Fig.~1d).

The equilibrium density of the Bose-Einstein condensate can be easily found with using the conservation of electric charge,  $n={n}_0+2{p}_0$. Here $n$ is the electron density corresponding to the initially instable system of normal electrons with the Fermi energy $\varepsilon_F>\varepsilon_0/2$, whereas ${n}_0$ and ${p}_0$ are the densities of normal electrons and coupled electrons, respectively, which correspond to the equilibrium state of the system with the Fermi energy of normal electrons ${\varepsilon}_{F0}=\varepsilon_0/2$ and the chemical potential of coupled electrons $\mu_0=\varepsilon_0$. Particularly, in the case of $\varepsilon_0>\Delta_0$, the equilibrium density of the Bose-Einstein condensate is
\begin{equation}\label{p0}
{p}_0=\frac{(m_1+m_2)\Delta\varepsilon}{2\pi\hbar^2},
\end{equation}
where the energy difference $\Delta\varepsilon=\varepsilon_F-\varepsilon_0/2$ has the physical meaning of the energy gain of the total system per one produced pair, which corresponds to the transition of the system from the nonequilibrium state pictured in Fig.~1b to the equilibrium state pictured in Fig.~1d. Keeping in mind that the present theory is developed under the assumption $p_0/n\ll1$, one can easily calculate the percentage of paired electrons $p_0/n$ for any quantum well with using Eq.~(\ref{p0}).
To find the applicability limits of Eq.~(\ref{p0}), it should be noted that the energy gain must satisfy the condition $\Delta\varepsilon\gg\Gamma_0$ in order to protect stability of the coupled electrons. It should be noted also that the overlap of the wave functions of neighbour electron pairs, $\varphi_0(\mathbf{r}_{12})$, must be negligible small in order to apply the model of ideal Bose gas to them as a first approximation. Therefore, the density (\ref{p0}) should be small enough to satisfy the condition $p_0r_0^2\ll1$, where $r_0$ is the size of the pair (see Fig.~1a). These two conditions can be satisfied simultaneously if $\Gamma_0\ll\hbar^2/(m_1+m_2)r_0^2$. Since this criterion conforms to the condition $\Gamma_0\ll\varepsilon_0-U_0(0)$ used above to derive the probability (\ref{Wks}), the present theory is self-consistent.

It follows from the aforesaid that the circularly polarized irradiation of 2D structures can produce the hybrid Bose-Fermi system consisting of the Fermi gas of normal conduction electrons and the Bose gas of coupled electrons. To analyze elementary excitations in the system, we have to take into account that the Fermi component changes the interaction between Bose particles, which is responsible for the dispersion of collective modes in the Bose-Einstein condensate, $\omega_\mathbf{K}$, where $\mathbf{K}$ is the wave vector of the mode. Generally, the frequency of the mode, $\omega_\mathbf{K}$, is defined by the conventional condition $\epsilon(\omega,\mathbf{K})=0$, where $\epsilon(\omega,\mathbf{K})=1-g_\mathbf{K}\Pi(\omega,\mathbf{K})$ is the dielectric function,
$\Pi(\omega,\mathbf{K})=\sum_{\bold{q}}[{{{p_{\bold{q}}-p_{\bold{q}+\bold{K}}}}}]/{{{[\hbar\omega+\varepsilon_0(\mathbf{q})-\varepsilon_0(\mathbf{q}+\mathbf{K})]}}}$
is the polarization operator  and $g_{\mathbf{K}}$ is the Fourier transform of the interaction between the Bose particles~\cite{PitStr}. Since a small number of electron pairs ($p_0r_0^2\ll1$) is immersed into the Fermi sea of conduction electrons, their Coulomb interaction is screened by many normal electrons and, therefore, can be described by the Fourier transform $g_{\mathbf{K}}=8\pi e^2/(K+2/r_s)$, where $r_s=\epsilon_0\hbar^2/(m_1+m_2)e^2$ is the effective screening radius assumed to satisfy the condition $r_s/r_0\gg1$, and $\epsilon_0$ is the static dielectric constant. Substituting the single-pair energy spectrum $\varepsilon_0(\mathbf{K})=\varepsilon_0+\hbar^2K^2/2(m_1+m_2)$ and the condensate density (\ref{p0}) into the polarization operator, we arrive at the energy spectrum of the collective mode,
$\hbar\omega_{\mathbf{K}}=\sqrt{2p_0g_\mathbf{K}E_\mathbf{K}+E^2_\mathbf{K}}$,
where $E_\mathbf{K}=\hbar^2K^2/2(m_1+m_2)$ is the kinetic energy of the composite boson. As expected, this collective mode has the sound-like dispersion $\omega_\mathbf{K}=v_sK$ for small wave vectors $K\ll1/r_s$, where $v_s=\sqrt{p_0g_{0}/(m_1+m_2)}$ is the sound velocity. Therefore, the considered Bose-Einstein condensate is superfluid if the velocity of its flow, $v$, satisfies the Landau criterion $v<v_s$.

To finalize the Letter, the applicability conditions of the dressed potential (\ref{U0}) to describe the electron pairing should be formulated. For the applicability of Eq.~(\ref{U0}), the field frequency, $\omega_0$, must satisfy the two conditions.
Firstly, $\omega_0\tau_e\gg1$,
where $\tau_e$ is the mean free time of charge carriers.
Secondly, the field frequency $\omega_0$ lies far from resonant frequencies corresponding to the optical transitions between different states of the paired charge carriers. Under the first condition, scattering
processes cannot destroy the paired carriers, whereas the second
condition allows to neglect the effect of the oscillating terms omitted in the dressed potential
(\ref{U0}) on the pairing (see the detailed discussion of the applicability limits of the dressed potential approach in Refs.~\cite{Kibis_2019,Kibis_2020,Iorsh_2020}). Among various 2D nanostructures containing charge carriers with different effective masses, quantum wells based on hole semiconductors look perspective to observe the discussed effects. Since the mobility of charge carriers in modern semiconductor quantum wells is of
$10^6-10^7\,\mathrm{cm}^2/\mathrm{V}\cdot\mathrm{s}$,
we have $\omega_0\tau_e\sim10$ near the high-frequency boarder of the microwave range,
$\nu_0=\omega_0/2\pi=100\,\mathrm{GHz}$. Normally, such a microwave frequency lies far from the resonant frequencies of paired electrons as well. Therefore, the two above-mentioned applicability conditions can be met simultaneously in quantum wells. Assuming that the reduced effective masses are $\overline{m},m\sim0.1m_e$, we arrive at the pair radius $r_0\sim10$~nm, the ground energy of coupled charge carriers $\varepsilon_0$ around $100$~meV and the energy broadening $\Gamma_0$ of meV scale for the relatively weak irradiation intensity $I\sim$~W/cm$^2$ with the frequency $\nu_0=\omega_0/2\pi=100\,\mathrm{GHz}$ (see plots in Fig.~2).  Correspondingly, the electron pairing can take place for the Fermi energy $\varepsilon_F$ of tens meV, which is attainable in modern semiconductor quantum wells.

{\it Conclusion.} We demonstrated theoretically that a circularly polarized irradiation can turn a 2D system containing degenerate electron gas with different effective masses into the hybrid Bose-Fermi system consisting of the Fermi gas of normal conduction electrons and the Bose gas of electron pairs (composite bosons) coupled by the irradiation. The found conditions of the electron pairing can be realized in modern 2D nanostructures and its possible manifestations, including the Bose-Einstein condensation of the pairs, can be observed in state-of-the-art measurements.

{\bf Funding.}
The reported study was funded by the Russian Science Foundation
(project 20-12-00001).

{\bf Disclosures.}
The authors declare no conflicts of interest.

{\bf Data Availability.}
No data were generated or analyzed in the presented research.

{\bf Supplemental Document.}
See Supplement 1 for supporting content.

\section{Supplement 1:\\Derivation of the decay probability of coupled electrons}

\setcounter{equation}{0}
\renewcommand{\theequation}{S\arabic{equation}}

The tunnel decay of the discussed electron pair corresponds to the transition $\varphi_\mathbf{K}\rightarrow\psi_{\mathbf{q},\mathbf{K}}$, which depends only on the relative momentum $\mathbf{q}$ and is pictured schematically in Fig.~1a (see the main text of the Letter).
To find the probability of this transition, let us construct the two-electron tunnel Hamiltonian. In the absence of the tunneling, the wave functions of the coupled two electrons, $\varphi_\mathbf{K}$, and the decoupled two electrons, $\psi_{\mathbf{q},\mathbf{K}}$, satisfy the orthogonality condition since they are separated by the impermeable potential barrier and, correspondingly, overlap of them is zero. Therefore, we can use them as a basis of the tunnel Hamiltonian which reads
\begin{equation}\label{H}
\hat{\cal H}=|\varphi_{0}\rangle\varepsilon_{0}\langle
\varphi_{0}|+\sum_{\mathbf{q}}|\mathbf{q}\rangle\varepsilon_{q}\langle
\mathbf{q}|
+\sum_\mathbf{q}\left[\,|\mathbf{q}\rangle T_\mathbf{q}\langle
\varphi_0|+\mathrm{H.c.}\right],
\end{equation}
where $|\varphi_{0}\rangle=\varphi_0(\mathbf{r}_{12})$  is the state of coupled electron pair with the energy $\varepsilon_0$, $|\mathbf{q}\rangle=(1/\sqrt{S})e^{i\mathbf{q}\mathbf{r}_{12}}$ is the state of decoupled electron pair with the relative momentum of the decoupled electrons $\mathbf{q}$ and the energy of their relative motion $\varepsilon_q=\hbar^2q^2/2m$, $S$ is the area of the considered 2D system, and the last term couples these  states with the tunnel matrix elements $T_\mathbf{q}=\langle\mathbf{q}|\hat{\cal H}|\varphi_0\rangle$.
The wave function satisfying the Schr\"odinger equation with the
Hamiltonian (\ref{H}) can be written as
$|\Psi\rangle=a_0(t)e^{-i\varepsilon_0t/\hbar}|\varphi_0\rangle+\sum_\mathbf{q}a_\mathbf{q}(t)e^{-i\varepsilon_{q}t/\hbar}|
{\mathbf{q}}\rangle$.  Substituting
this wave function into the Schr\"odinger equation with the Hamiltonian (\ref{H}),
$i\hbar\partial_t|\Psi\rangle=\hat{\cal H}|\Psi\rangle$, we arrive at the
equations for the expansion coefficients,
\begin{eqnarray}
i\hbar\dot{a}_0(t)&=&\sum_\mathbf{q}e^{i(\varepsilon_0-\varepsilon_{q})t/\hbar}
T^\ast_\mathbf{q}a_\mathbf{q}(t),\label{dyn1}\\
i\hbar\dot{a}_\mathbf{q}(t)&=&e^{i(\varepsilon_{q}-\varepsilon_0)t/\hbar}T_\mathbf{q}a_s(t).\label{dyn2}
\end{eqnarray}
Let an electron pair be in the coupled state at the time $t=0$, i.e.
$a_0(0)=1$ and $a_\mathbf{q}(0)=0$. Then the integration of
Eq.~(\ref{dyn2}) results in
\begin{equation}\label{b}
a_\mathbf{q}(t)=-\frac{iT_\mathbf{q}}{\hbar}\int_0^{\,t}e^{i(\varepsilon_{q}-\varepsilon_0)t^\prime/\hbar}a_0(t^\prime)dt^\prime.
\end{equation}
Since the considered system is axially symmetrical, the matrix
element $T_\mathbf{q}$ depends only on the electron energy,
$\varepsilon_{q}$ and, therefore, can be denoted as
$T_\mathbf{q}=T_{\varepsilon_{q}}$. Substituting Eq.~(\ref{b})
into Eq.~(\ref{dyn1}), we arrive at the expression
\begin{align}\label{a}
&\dot{a}_0(t)=\nonumber\\
&-\frac{Sm}{2\pi\hbar^4}\int_0^{\,\infty}
d\varepsilon_{q}\,\,|T_{\varepsilon_{q}}|^2\int_0^{\,t}e^{i(\varepsilon_0-\varepsilon_{q})(t-t^\prime)/\hbar}a_0(t^\prime)dt^\prime.
\end{align}
This is still an exact equation since we just replaced two
differential equations (\ref{dyn1})--(\ref{dyn2}) with one linear
differential-integral equation (\ref{a}). Next, we make the
approximation. Namely, let us assume that the tunneling
between the states $|\varphi_0\rangle$ and $|\mathbf{q}\rangle$ is weak to satisfy the condition $\Gamma_0\ll\varepsilon_0-U_0(0)$, where $\Gamma_0$ is the tunnel-induced broadening of the coupled state energy $\varepsilon_0$. Then the quantity $|T_{\varepsilon_{q}}|^2$ varies little
around $\varepsilon_{q}=\varepsilon_0$ for which the time integral
in Eq.~(\ref{a}) is not negligible. Therefore, the energy of the decoupled electron pair, $\varepsilon_{q}$, is
near the energy of the coupled pair, $\varepsilon_0$.
As a consequence, one can make the replacement
$|T_{\varepsilon_{q}}|\rightarrow |T_{\varepsilon_0}|$ and replace
the lower limit in the $\varepsilon_{q}$ integration with
$-\infty$. As a result, it follows
from Eq.~(\ref{a}) that $a_0(t)=e^{-\Gamma_0 t/2\hbar}$, where $\Gamma_0=2\pi\sum_{\mathbf{q}}|T_{\mathbf{q}}|^2\delta(\varepsilon_0-\varepsilon_{q})$.
Substituting the found amplitude $a_0(t)$ into Eq.~(\ref{b}), the amplitude of the tunnel transition $\varphi_\mathbf{K}\rightarrow\psi_{\mathbf{q},\mathbf{K}}$ during time $t$ reads
\begin{equation}\label{bk}
a_\mathbf{q}(t)=-T_{\varepsilon_{0}}\frac{e^{i(\varepsilon_{q}-\varepsilon_0)t/\hbar-\Gamma_0
t/2\hbar}-1}{\varepsilon_q-\varepsilon_{0}+i\Gamma_0/2}.
\end{equation}
Correspondingly, the sought probability of tunnel decay of the two coupled electrons into the state of two decoupled electrons with the relative momentum $\mathbf{q}$ is
\begin{equation}\label{Wkss}
W_{\mathbf{q}}=|a_{\mathbf{q}}(\infty)|^2=\frac{\hbar^2\Gamma_0}{Sm[(\varepsilon_q-\varepsilon_{0})^2+(\Gamma_0/2)^2]}.
\end{equation}


\begin{thebibliography}{99}

\bibitem{Oka_2019}
T. Oka and S. Kitamura, Annu. Rev. Condens. Matter. Phys. {\bf 10}, 387 (2019).

\bibitem{Bukov_2015}
M. Bukov, L. D'Alessio, and A. Polkovnikov, Adv. Phys. {\bf
64}, 139 (2015).

\bibitem{Lindner_2011}
N. H. Lindner, G. Refael and V. Galitski, Nat. Phys. {\bf 7}, 490
(2011).

\bibitem{Savenko_2012}
I. G. Savenko, O. V. Kibis and I. A. Shelykh, Phys. Rev. A {\bf 85}, 053818 (2012).

\bibitem{Kozin_2018}
V. K. Kozin, I. V. Iorsh, O. V. Kibis and I. A. Shelykh, Phys. Rev. B {\bf 97}, 035416
(2018).

\bibitem{Rechtsman_2013}
M. C. Rechtsman, J. M. Zeuner, Y. Plotnik, Y. Lumer, D. Podolsky,
F. Dreisow, S. Nolte, M. Segev and A. Szameit, Nature {\bf 496}, 196 (2013).

\bibitem{Kibis_2019}
O. V. Kibis, Phys. Rev. B {\bf 99}, 235416 (2019).

\bibitem{Kibis_2020}
O. V. Kibis, M. V. Boev, and V. M. Kovalev, Phys. Rev. B {\bf 102}, 075412 (2020).

\bibitem{Iorsh_2020}
O. V. Kibis, S. A. Kolodny, and I. V. Iorsh, Opt. Lett. {\bf 46}, 50 (2021).

\bibitem{Kavokin_2005}
A. V. Kavokin, I. A. Shelykh and M. M. Glazov, Phys. Stat. Sol. (c) {\bf 2}, 914 (2005).

\bibitem{Kroo_2014}
N. Kro\'o, P. R\'acz and S. Varr\'o, EPL {\bf 105}, 67003 (2014).

\bibitem{Landau_3}
L. D. Landau and E. M. Lifshitz, {\it Quantum mechanics:
Non-relativistic theory} (Pergamon Press, Oxford, 1965).

\bibitem{Landau_5}
L. D. Landau and E. M. Lifshitz, {\it Statistical physics, Part 1} (Pergamon Press, Oxford, 1980).

\bibitem{PitStr}
L. Pitaevskii and S. Stringari, \textit{Bose-Einstein condensateion} (Clarendon Press, Oxford, 2003).

\end{thebibliography}
\end{document}